\documentclass[12pt]{article}
\usepackage{amsmath}
\usepackage{graphicx,psfrag,epsf}
\usepackage{subfigure,amsmath,latexsym,amssymb}
\usepackage{float,epsfig,multirow,rotating,times}
\usepackage{upgreek,wrapfig}
\usepackage{enumerate}
\usepackage{natbib}

\newcommand{\blind}{0}

\newcommand {\ctn}{\citet}

\newcommand{\topline}{\hrule height 1pt width \textwidth \vspace*{2pt}}
\newcommand{\botline}{\vspace*{2pt}\hrule height 1pt width \textwidth \vspace*{4pt}}
\newtheorem{algo}{Algorithm}
\numberwithin{algo}{section}

\newcommand{\btheta}{\boldsymbol{\theta}}

\newcommand{\bphi}{\boldsymbol{\phi}}
\newcommand{\bPhi}{\boldsymbol{\Phi}}

\newcommand{\bxi}{\boldsymbol{\xi}}

\newcommand{\bTheta}{\boldsymbol{\Theta}}
\newcommand{\bgamma}{\boldsymbol{\gamma}}

\newcommand{\bzeta}{\boldsymbol{\zeta}}

\newcommand{\bC}{\boldsymbol{C}}

\newcommand{\bG}{\boldsymbol{G}}

\newcommand{\bU}{\boldsymbol{U}}

\newcommand{\by}{\boldsymbol{y}}
\newcommand{\bY}{\boldsymbol{Y}}

\begin{document}

\bibliographystyle{natbib}

\def\spacingset#1{\renewcommand{\baselinestretch}%
{#1}\small\normalsize} \spacingset{1}


\if0\blind
{
  \title{\bf Clustering Categorical Time Series into Unknown Number of Clusters:
  A Perfect Simulation based Approach}
  \author{Sabyasachi Mukhopadhyay\\
  Southampton Statistical Sciences Research Institute\\
  University of Southampton\\ 
  Southampton, UK\\
  and\\
  Sourabh Bhattacharya\thanks{Corresponding e-mail: sourabh@isical.ac.in}\\
  Bayesian and Interdisciplinary Research Unit\\
  Indian Statistical Institute
}
\maketitle
} \fi
\if1\blind
{
 \bigskip
 \bigskip
 \bigskip
 \begin{center}
 {\LARGE\bf Clustering Categorical Time Series into Unknown Number of CLusters:
 A Perfect Simulation based Approach}
 \end{center}
 \medskip
} \fi

\bigskip

\begin{abstract}

\ctn{Pamminger10} considered a Bayesian approach to model-based clustering of categorical time series assuming
a fixed number of clusters. But the popular methods for selecting the number of clusters, for example,
the Bayes Information Criterion (BIC), turned out to have severe problems in the categorical time series
context. 

In this paper, we circumvent the difficulties of choosing the number of clusters by 
adopting the Bayesian semiparametric mixture model approach introduced by \ctn{Bhattacharya08},
who assume that the number of clusters is a random quantity, but is bounded above by a (possibly large)
number of clusters.  
We adopt the perfect simulation approach of \ctn{Sabya10b} for posterior
simulation for completely solving the problems of convergence of the underlying Markov chain Monte
Carlo (MCMC) approach. 

Importantly, within our main perfect simulation algorithm, there arose the necessity
to simulate perfectly from the joint distribution of a set of continuous random variables
with log-concave full conditional densities. We propose and develop a novel and efficient perfect simulation methodology
for joint distributions with log-concave full conditionals. This
perfect sampling methodology is of independent interest as well since in a very 
large and important class of Bayesian
applications the full conditionals turn out to be log-concave. 

We will consider application of our model and methodology to the Austrian wage mobility data, also
analysed by \ctn{Pamminger10}, and adopting the methods developed in \ctn{Sabya10a}, \ctn{Sabya10},
will obtain the posterior modes of clusterings and also the desired highest posterior distribution credible
regions of the posterior distribution of clusterings. With these summaries of the posterior
distribution of clustering we will detail the consequences of ignoring uncertainty in the number
of clusters in the approach of \ctn{Pamminger10}.

\end{abstract}
\noindent%
{\it Keywords:}
Bounding chains; Categorical time series; Dirichlet process; Gibbs sampling; Mixtures; Optimization; Perfect Sampling

\spacingset{1.45}

\section{Introduction}
\label{sec:introduction}

We consider the problem of clustering a panel of categorical time series $\by_i;i=1,\ldots,N$
into several classes (components), assuming that the number of classes is unknown.
The known number of components situation has been recently handled by \ctn{Pamminger10}, who
consider a Bayesian mixture model based approach with a fixed number of components.
However, the authors reported serious difficulties in reliably determining the appropriate
number of components using the traditional approaches like Bayes Information Criterion (BIC).

We completely avoid the difficulties of the fixed components approach by adopting the
assuming that the number of components is unknown, but is bounded above by a number
specified by experts, the upper bound signifying that the number of possible clusters
of the time series can not exceed the specified upper limit. Such a model has been
proposed by \ctn{Bhattacharya08}; see also \ctn{Sabya10} and \ctn{Sabya10a}.
We develop a perfect simulation method for sampling exactly from the underlying posterior
distribution. Perfect simulation for mixtures with unknown number of components has been
developed by \ctn{Sabya10b}, but in the time series context there are some additional complications,
to be explained in due course. 

Indeed, these additional difficulties led us to develop
a general perfect simulation methodology in the case of joint distributions with log-concave full conditional
distributions, which is of independent interest.

With our new developments related to pefect sampling, we then proceed to analyze the Austrian
wage mobility data, obtaining the modes of the posterior distribution of clustering as well
as the desired highest posterior distribution credible regions, using the methods detailed
in \ctn{Sabya10a} and \ctn{Sabya10}. In particular, we demonstrate that ignoring uncertainty
in the number of clusters in the approach of \ctn{Pamminger10} seriously affects inference.

The rest of our paper is structured as follows. In Section \ref{sec:categorical_mixture},
adopting the mixture model of \ctn{Bhattacharya08} we model the time series as mixtures of 
unknown number of components, and in Section \ref{sec:fullcond} provide the full conditional
distributions to be used for perfect simulation, along with the need for perfect simulation
from joint distributions of continuous parameters with log-concave full conditionals in our problem.
Some more details are presented in the supplement, the sections of which we refer to by 
using the prefix ``S-".
In Section \ref{sec:perfect_logconcave} we introduce our perfect simulation idea in the case
of continuous joint distributions having log-concave full conditionals. 
Using this development, and adopting the perfect simulation idea for mixtures with unknown 
number of components proposed by \ctn{Sabya10b} we then present the relevant perfect 
simulation methodology for our categorical time series problem. 

\section{Mixtures of categorical time series with unknown number of components}
\label{sec:categorical_mixture}
In this work we confine ourselves to mixtures of categorical time series with Markov chain
clustering, which has also been the main aspect of study in \ctn{Pamminger10}, albeit
the latter consider only fixed number of components. In what follows we shall borrow
some notation already described in \ctn{Pamminger10}.

We consider the mixture model of the following form:
for $i=1,\ldots,N$,
\begin{equation}
f(\by_i\mid\bTheta)=\frac{1}{M}\sum_{h=1}^M\prod_{r=1}^{T_i}f(y_{ir}\mid y_{i,r-1},\btheta_h)
=\frac{1}{M}\sum_{h=1}^M\prod_{s=1}^K\prod_{t=1}^K\theta^{N_{i,st}}_{h,st},
\label{eq:categorical_mixture}
\end{equation}
where $N_{i,st}=\#\{y_{ir}=t,y_{i,r-1}=s\}$ is the number of transitions from state $s$ to state $t$
observed in time series $i$, and, for each $h\in\{1,\ldots,M\}$, $\btheta_h=((\theta_{h,st}));s,t=1,\ldots,K$,
is the transition matrix of the underlying Markov chain model consisting of $K$ states.
The latter satisfies $\sum_{t=1}^K\theta_{h,st}=1$ $\forall h,s$.
In (\ref{eq:categorical_mixture}), $M$ is the maximum number of components, specified, perhaps,
by some expert; however, \ctn{Sabya11} show how $M$ can be obtained objectively and optimally from a
Bayesian asymptotics perspective.

We next consider the following Dirichlet process (DP) prior for $\bTheta$:
for $h=1,\ldots,M$,
\begin{align}
\btheta_h&\stackrel{iid}{\sim}\bG\label{eq:dp1}\\
\bG&\sim DP(\alpha\bG_0)\label{eq:dp2}
\end{align}
Under $\bG_0$, for $h=1,\ldots,M$ and $s=1,\ldots,K$,
\begin{align}
(\theta_{h,s1},\ldots,\theta_{h,sK})&\sim Dirichlet(\gamma_{s1},\ldots,\gamma_{sK})
\label{eq:dp3}
\end{align}
In addition, we assume that
\begin{align}
\gamma_{st}&\sim Gamma(a_{st},b_{st});\ \ \ s=1,\ldots,K; t=1,\ldots,K, 
\label{eq:gamma_prior}
\end{align}
where $Gamma(a,b)$ denotes the gamma distribution of the form
$\frac{b^a}{\Gamma(a)}x^{a-1}\exp(-bx)$, having mean $a/b$ and variance $a/b^2$.
We remark that \ctn{Pamminger10} assumed a discrete prior distribution on $\{\gamma_{st};s,t=1,\ldots,K\}$,
namely, the negative multinomial distribution. However, continous priors, such as ours,
are perhaps more appropriate and more natural.


The Dirichlet process prior assumption entails that mixture
model (\ref{eq:categorical_mixture}) reduces to the following form:
\begin{equation}
f(\by_i\mid\bTheta_M)=\sum_{\ell=1}^p\pi_{\ell}\prod_{r=1}^{T_i}f(y_{ir}\mid y_{i,r-1},\bphi_{\ell})
=\sum_{\ell=1}^p\pi_{\ell}\prod_{s=1}^K\prod_{t=1}^K\phi^{N_{i,st}}_{\ell,st},
\label{eq:categorical_mixture2}
\end{equation}
where $\bphi_{\ell}$ denotes the $\ell$-th distinct component among $\bTheta_M=\{\btheta_1,\ldots,\btheta_M\}$,
and $\pi_{\ell}=M_{\ell}/M$, with $M_{\ell}=\#\{h:\btheta_h=\bphi_{\ell}\}$. In (\ref{eq:categorical_mixture2}),
$p~(1\leq p\leq M)$ denotes the {\it random} number of distinct mixture components.

\section{Full conditional distributions}
\label{sec:fullcond}

Let $\bY=\{\by_1,\ldots,\by_n\}$ denote the data set.
We define the set of allocation variables $Z=\{z_1,\ldots,z_n\}$, where $z_i=j$ if $\by_i$
arises from the $j$-th component.
Letting $\bPhi=\{\bphi_1,\ldots,\bphi_k\}$ denote the distinct components in $\Theta_M$,
the element $c_j$ of the configuration vector $C=\{c_1,\ldots,c_M\}$ is defined as $c_j=\ell$ if and only if
$\btheta_j=\bphi_{\ell}$; $j=1,\ldots,M$, $\ell=1,\ldots,k$. Thus, $(Z,\Theta_M)$ is reparameterized
to $(Z,C,k,\bPhi)$, $k$ denoting the number of distinct components in $\Theta_M$.

\subsection{Full conditionals of $\{z_1,\ldots,z_n\}$}
\label{subsec:fullcond_z}

For $i=1,\ldots,n$, let $Z_{-i}=\{z_1,\ldots,z_{i-1},z_{i+1},\ldots,z_n\}$, and
let $C$ consist of $k$ distinct components. Then, denoting the set $\{\gamma_{st};s,t=1,\ldots,K\}$
by $\bgamma$, the full conditional distribution of $z_i$ is given by
\begin{align}
[z_i=r\mid \bY, Z_{-i},C,\bPhi,\bgamma,k]&\propto 
\prod_{s=1}^K\prod_{t=1}^K\theta^{N_{i,st}}_{r,st}
\label{eq:fullcond_z}
\end{align}

\subsection{Full conditionals of $\{c_1,\ldots,c_M\}$}
\label{subsec:fullcond_c}

To obtain the full conditional of $c_r;r=1,\ldots,M$, first let $k_{r}$ denote the number of distinct values in
$\bTheta_{-rM}=\{\btheta_1,\ldots,\btheta_{r-1},\btheta_{r+1},\ldots,\btheta_M\}$, and let
$\bphi^{(r^*)}_{\ell}$; $\ell=1,\ldots,k_{r}$ denote
the distinct values.
Also suppose that $\bphi^{(r^*)}_{\ell}$ occurs $M_{\ell r}$ times.
Then the conditional distribution of $c_r$ is given by
\begin{equation}
[c_r=\ell\mid \bY,Z,C_{-r},\bPhi,\bgamma,k_r]=\left\{\begin{array}{c}\kappa q_{\ell r}\hspace{2mm}\mbox{if}\hspace{2mm}
\ell=1,\ldots,k_r\\ \kappa q_{0r}\hspace{2mm}\mbox{if}\hspace{2mm}\ell=k_r+1\end{array}\right.
\label{eq:fullcond_c}
\end{equation}
where
\begin{align}
q_{\ell r}&\propto M_{\ell r}\times
\prod_{s=1}^K\prod_{t=1}^K\phi^{\sum_{i:z_i=r}N_{i,st}}_{\ell,st}
\label{eq:fullcond_c1}
\end{align}
and
\begin{align}
q_{0 r}&\propto \alpha\times
\prod_{s=1}^K
\frac{\Gamma\left(\sum_{t=1}^K\gamma_{st}\right)}
{\prod_{t=1}^K\Gamma\left(\gamma_{st}\right)}
\times\prod_{s=1}^K
\frac{\prod_{t=1}^K\Gamma\left(\sum_{i:z_i=r}N_{i,st}+\gamma_{st}\right)}
{\Gamma\left(\sum_{t=1}^K\sum_{i:z_i=r}N_{i,st}+\sum_{t=1}^K\gamma_{st}\right)}
\label{eq:fullcond_c2}
\end{align}

\subsection{Full conditionals of $\{\bphi_{\ell};\ell=1,\ldots,k\}$}
\label{subsec:fullcond_phi}

Assuming that there are $k$ distinct components in $C$, the full conditional distribution
of $\bphi_{\ell};\ell=1,\ldots,k$, is given by
\begin{align}
[\bphi_{\ell}\mid \bY, Z, C,\bPhi_{-\ell},\bgamma,k]
&= \prod_{s=1}^K\prod_{t=1}^K\phi^{\sum_{i:z_i=j}\sum_{j:c_j=\ell}N_{i,st}+\gamma_{st}-1}_{\ell,st}\notag\\
& \ \ \ \ \times\prod_{s=1}^K\frac{\Gamma\left(\sum_{t=1}^K\sum_{i:z_i=j}\sum_{j:c_j=\ell}N_{i,st}+\sum_{t=1}^K\gamma_{st}\right)}
{\prod_{t=1}^K\Gamma\left(\sum_{i:z_i=j}\sum_{j:c_j=\ell}N_{i,st}+\gamma_{st}\right)},
\label{eq:fullcond_phi}
\end{align}
which are conditionally independent of $\bPhi_{-\ell}$.

The conditional mean and variance of $\phi_{\ell,s^*t^*}$ are given, respectively, by
\begin{equation}
\zeta_{\ell,s^*t^*}=E\left[\phi_{\ell,s^*t^*}\mid \bY, Z, C,\bgamma,k\right]=
\frac{\gamma_{s^*t^*}+\sum_{i:z_i=j;j:c_j=\ell} N_{i,s^*t^*}}
{\sum_{t=1}^K\left(\gamma_{s^*t}+\sum_{i:z_i=j;j:c_j=\ell} N_{i,s^*t}\right)},
\label{eq:phi4}
\end{equation}
and
\\[2mm]
$\varphi_{\ell,s^*t^*}=Var\left[\phi_{\ell,s^*t^*}\mid  \bY, Z, C,\bgamma,k\right]$
\begin{equation}
=\frac{\left(\gamma_{s^*t^*}+\sum_{i:z_i=j;j:c_j=\ell} N_{i,s^*t^*}\right)
\left\{\sum_{t\neq t^*}\left (\gamma_{s^*t}+\sum_{i:z_i=j;j:c_j=\ell} N_{i,s^*t}\right)\right\}}
{\left\{\sum_{t=1}^K\left(\gamma_{s^*t}+\sum_{i:z_i=j;j:c_j=\ell} N_{i,s^*t}\right)\right\}^2
\left\{1+\sum_{t=1}^K\left(\gamma_{s^*t}+\sum_{i:z_i=j;j:c_j=\ell} N_{i,s^*t}\right)\right\}},
\label{eq:phi5}
\end{equation}
It follows that 
\begin{equation}
\frac{\varphi_{\ell,s^*t^*}}
{\zeta_{\ell,s^*t^*}
\left(1-\zeta_{\ell,s^*t^*}\right)}
=\frac{1}{1+\sum_{t=1}^K\left(\gamma_{s^*t}+\sum_{i:z_i=j;j:c_j=\ell} N_{i,s^*t}\right)}.
\label{eq:cluster1}
\end{equation}
Accordingly, as in \ctn{Pamminger10}, but somewhat differently, we can interpret 
$\Sigma_{\ell,s^*}=$$\sum_{t=1}^K(\gamma_{s^*t}$$+\sum_{i:z_i=j;j:c_j=\ell} N_{i,s^*t})$
as a {\it conditional} measure of heterogeneity in the corresponding rows of $\bphi_{\ell}$
of the $\ell$-th cluster. Small values of $\Sigma_{\ell,s^*}$ implies high degree of variability
of the individual transition probabilities $\phi_{\ell,s^*t^*}$ and large deviations of
$\bphi_{\ell,s^*}=\left(\phi_{\ell,s^*1},\ldots,\phi_{\ell,s^*K}\right)$ from the group mean
$\bzeta_{\ell,s^*}=\left(\zeta_{\ell,s^*1},\ldots,\zeta_{\ell,s^*K}\right)$.
Large values of $\Sigma_{\ell,s^*}$ indicate small variability in the $s^*$-th row, implying
that the individual transition probabilities $\phi_{\ell,s^*t^*}$ are nearly the same as
as the group means $\zeta_{\ell,s^*t^*}$. 

Interestingly, for the purpose of perfect simulation, the full or marginal conditional
distributions of $\phi_{\ell,s^*t^*}$, given below, will be shown to be more
important than those of $\bphi_{\ell}$, 
even though the latter is just the standard Dirichlet distribution and straightforward to simulate from.

\subsubsection{Full and marginal conditionals of $\phi_{\ell,s^*t^*}$}
The full conditional of $\phi_{\ell,s^*t^*}$ is given by
\\[2mm]
$[\phi_{\ell,s^*t^*}\mid \bY, Z, C,\bPhi_{-\ell,-s^*,-t^*},\bgamma,k]$
\begin{align}
&\propto \phi^{\sum_{i:z_i=j}\sum_{j:c_j=\ell}N_{i,s^*t^*}+\gamma_{s^*t^*}-1}_{\ell,s^*t^*}\notag\\
&\ \ \ \ \times\left(1-\sum_{t=1}^K\phi_{\ell,s^*t}\right)^{\sum_{i:z_i=j}\sum_{j:c_j=\ell}N_{i,s^*K}+\gamma_{s^*K}-1}\notag\\
\label{eq:fullcond_phi2}
\end{align}
In the above, $\bPhi_{-\ell,-s^*,-t^*}$ denotes $\bPhi$ without $\phi_{\ell,s^*t^*}$.

The marginal conditional of $\phi_{\ell,s^*t^*}$ is given by
\\[2mm]
$[\phi_{\ell,s^*t^*}\mid \bY, Z, C,\bPhi_{-\ell,-s^*,-t^*},\bgamma,k]$
\begin{align}
&\propto \phi^{\sum_{i:z_i=j;j:c_j=\ell}N_{i,s^*t^*}+\gamma_{s^*t^*}-1}_{\ell,s^*t^*}\notag\\
&\ \ \ \ \times\left(1-\phi_{\ell,s^*t^*}\right)^{\sum_{t\neq t^*}
\left(\sum_{i:z_i=j;j:c_j=\ell}N_{i,s^*t}+\gamma_{s^*t}\right)-1},
\label{eq:marginal_cond_phi2}
\end{align}
which is a $Beta$ distribution with parameters $\sum_{i:z_i=j;j:c_j=\ell}N_{i,s^*t^*}+\gamma_{s^*t^*}$
and $\sum_{t\neq t^*}\left(\sum_{i:z_i=j;j:c_j=\ell}N_{i,s^*t}+\gamma_{s^*t}\right)$.

\subsection{Full conditionals of $\{\gamma_{st};s,t=1,\ldots,K\}$}
\label{subsec:fullcond_gamma}

Assuming that $C$ consists of $k$ distinct components, the full
conditional distribution of $\gamma_{s^*t^*}$, for
$s^*=1,\ldots,K$, and $t^*=1,\ldots,K$, is given by
\begin{align}
[\gamma_{\ell,s^*t^*}\mid \bY, Z,C,\bPhi,\bgamma_{-s^*,-t^*},k]&\propto 
\left(\prod_{\ell=1}^k\phi^{\gamma_{s^*t^*}-1}_{\ell,s^*t^*}\right)
\times 
\left(\frac{\Gamma\left(\sum_{t=1}^K\gamma_{s^*t}\right)}
{\Gamma\left(\gamma_{s^*t^*}\right)}\right)^k\notag\\
&\times \gamma^{a_{s^*t^*}-1}_{s^*t^*}\exp\left(-b_{s^*t^*}\gamma_{s^*t^*}\right)
\label{eq:fullcond_gamma}
\end{align}
In the above, $\bgamma_{-s^*,-t^*}$ denotes all elements of the $\bgamma$-parameters
except $\gamma_{s^*t^*}$.

 \subsection{Relabeling $C$}
 \label{subsec:relabeling}
 Simulation of $C$ by successively simulating from the full conditional distributions (\ref{eq:fullcond_c}) incurs
 a labeling problem. For instance, it is possible that all $c_j$ are equal even though each of them corresponds to
 a distinct $\btheta_j$. For an example, suppose that $\bPhi$ consists of $M$ distinct elements, and $c_j=M$ $\forall j$.
 Then although there are actually $M$ distinct components, one ends up obtaining just one distinct component.
 For perfect sampling \ctn{Sabya10b} created a labeling method which relabels $C$ such that the relabeled version,
 denoted by $S=(s_1,\ldots,s_M)'$, coalesces if $C$ coalesces.
 To  construct $S$ we first simulate $c_j$ from (\ref{eq:fullcond_c}); if $c_j\in\{1,\ldots,k_j\}$, then we set $\btheta_j=\bphi_{c_j}$
 and if $c_j=k_j+1$, we draw $\btheta_j=\bphi_{c_j}\sim G_j$. The elements of $S$ are obtained from
 the following definition of $s_j$:
 $s_j=\ell$ if and only if $\btheta_j=\bphi_{\ell}$. Note that $s_1=1$ and $1\leq s_j\leq s_{j-1}+1$.
 \ctn{Sabya10b} proved that coalescence of $C$ implies the coalescence of $S$, irrespective of the value of $\bPhi$.

\subsection{Full conditionals using $S$}
\label{subsec:fullcond_s}
With the introduction of $S$ it is now required to modify some of the full conditionals of the unknown random variables, in addition
to introduction of the full conditional distribution of $S$.
The form of the full conditional $[z_i\mid \bY,S,k,\bPhi,\bgamma]$ remains the same as
(\ref{eq:fullcond_z}), but $\Theta_M$ involved in the right hand side
of (\ref{eq:fullcond_z}) is now obtained from $S$ and $\bPhi$. The modified  full conditional of $c_j$, which we denote
by $[c_j\mid \bY,Z,S_{-j},k_j,\bPhi]$, now depends upon $S_{-j}$, rather than $C_{-j}$, the notation being clear from the context.
The form of this full conditional remains the same as (\ref{eq:fullcond_c}) but now the distinct components
$\bphi^{j^*}_{\ell}$; $\ell=1,\ldots,k_{j}$ are associated with the corresponding components of $S$ rather than $C$. The form of the modified full
conditional distribution of $\bphi_{\ell}$, which we now denote by $\left[\bphi_{\ell}\mid \bY,Z,S,k\right]$, remains the same as
(\ref{eq:fullcond_phi}), only $C$ must be replaced with $S$.
Also $k$ and $k_j$ are now assumed to be associated with $S$.
The conditional posterior $[S\mid \bY,C,\bPhi,\bgamma,k]$ gives point mass to $S^*$, where
$S^*=\{s^*_1,\ldots,s^*_M\}$ is the relabeling obtained from $C$ and $\Theta_M$
following the method described in Section \ref{subsec:relabeling}.

For the construction of bounds, the individual full conditionals $[s_j\mid Y, S_{-j},C,\bPhi,\bgamma,k]$,
giving full mass to $s^*_j$, will be considered due to
convenience of dealing with distribution functions of one variable.
It follows that once $Z$ and $C$ coalesces, $S$ and $\bPhi$ must also coalesce.
In the next section we describe how to construct efficient bounding chains for $Z$, $C$ and $S$.
Bounding chains for $S$ are not strictly necessary as it is possible to optimize the bounds for $Z$ and $C$ with respect to $S$, but the efficiency of the other bounding chains
is improved, leading to an improved perfect sampling algorithm,
if we also construct bounding chains for $S$.

\subsection{Need for perfect simulation of $(\bPhi,\bgamma)$ given the rest}
\label{subsec:perfect_need}

The perfect sampling methodology for mixtures of unknown number of components
developed in \ctn{Sabya10b} can be envisaged for simulating exactly from the posterior
in this categorical time series problem.
Their method requires simulation of the discrete parameters $(Z,C,S)$ only and not the continuous parameters
$\bPhi$ and $\bgamma$ until coalescence
of the discrete parameters. Simulation of the continuous parameters is necessary only after
the discrete parameters have coalesced. In our example, however, simulation of $\bPhi$ and $\bgamma$
given $Z$ and $C$, even after coalescence of the latter, is not straightforward. This is because
there does not seem to exist any method of directly simulating from the (joint) full conditional of $(\bPhi,\bgamma)$
and so it is required to simulate from the component-wise full conditionals of $\bphi_{\ell}$ given $\bgamma$, and
from the (non-standard) component-wise full conditionals of $\gamma_{st}$, given $Z$, $S$, and $\bgamma_{-s,-t}$ and $\bPhi$,
and although the initial values of $Z$ and $S$ are the coalesced values of the respective
bounding chains, the initial values of $\bgamma$ for simulating $\bphi_{\ell}$
or the initial values of $\bPhi$ and $\bgamma_{-s,-t}$ for simulating $\gamma_{st}$, are not available. The non-availability
of starting values is due to the fact that before coalescence of $(Z,C,S)$,
$\bPhi$ and $\bgamma$ are not simulated at all.

The above problem calls for the need for perfect simulation of $\bPhi$ and $\bgamma$ given $(Z,S)$, using the
full conditionals of $\bPhi_{\ell}$ and $\gamma_{st}$, given by (\ref{eq:fullcond_phi}) and
(\ref{eq:fullcond_gamma}), respectively. Thus, our main perfect
simulation methodology must proceed via incorporation of another perfect sampling method involving
the full conditionals of $\gamma_{st}$. But Gibbs sampling-based perfect simulation in the case
of continuous parameters is not developed in the literature. In this paper, we propose and develop
a novel and general perfect simulation methodology using full conditional distributions of continuous
parameters. All we require is that the full conditionals are log-concave. We then specialize our
general methodology to the problem of perfectly simulating $\bPhi$ and $\bgamma$ given $Z,S$ within the perfect
sampling methodology of \ctn{Sabya10b}.
However, as mentioned already, for perfect sampling, we shall need to utilize the full conditional
of $\phi_{\ell,st}$, given by (\ref{eq:fullcond_phi2}) rather than that of $\bphi_{\ell}$, given by
(\ref{eq:fullcond_phi}).
Indeed, it is easy to see that the full conditionals of $\phi_{\ell,st}$ and $\gamma_{\ell,st}$ satisfy
\begin{align}
&\frac{d^2}{d\phi^2_{\ell,st}}[\phi_{\ell,st}\mid\bY,Z,S,\bPhi_{-\ell,-s,-t},\bgamma,k]<0,\label{eq:lc1}\\
&\frac{d^2}{d\gamma^2_{st}}[\gamma_{st}\mid\bY,Z,S,\bPhi,\bgamma_{-s,-t},k]<0,\label{eq:lc2}
\end{align}
provided that $\sum_{i:z_i=j}\sum_{j:c_j=\ell}N_{i,st}+\gamma_{st}>1$ and $a_{st}>1$.
The proof of (\ref{eq:lc1}) follows by simple differentiation, while the proof of (\ref{eq:lc2})
also requires the formula (see \ctn{Bowman88}):
\begin{equation}
\frac{d^2}{dx^2}\log\left[\Gamma(x)\right]=\frac{1}{x}+\frac{1}{2x^2}
+\frac{2\pi}{x}\int_0^{\infty}\frac{y\sqrt{t}}{(x^2+t)(y-1)^2}dt,
\label{eq:lc3}
\end{equation}
where $y=\exp(2\pi\sqrt{t})$.
Using the above formula the proof of (\ref{eq:lc2}) follows in similar lines
as the proof of Proposition 2 of \ctn{He98}.
Note that although it is possible to integrate out $\bPhi$ thanks to conjugacy
and obtain the marginalized full conditionals of $(Z,C,\bgamma)$ (see Section \ref{sec:marginalized_fullcond}
of the supplement), it can be
easily verified that the resulting expression
for the full conditional of $\gamma_{st}$ need not admit log-concavity; 
see Section \ref{subsec:gamma_logconcavity} of the supplement for details.
This lack of log-concavity makes it difficult to generate perfect samples from the full conditional of $\bgamma$.

\section{Perfect simulation in posteriors with log-concave full conditionals}
\label{sec:perfect_logconcave}

Before introducing our perfect simulation idea
in Gibbs sampling for continuous, log-concave, full conditionals, we first provide a brief description of
adaptive rejection sampling (ARS) following \ctn{Gilks92a}.

\subsection{Overview of ARS}
\label{subsec:ars}

Assuming that it is required to sample from a log-concave density $g(\cdot)$ with
$g(\cdot)$ continuous and differentiable everywhere on a set $D$, let us suppose that
$h(x)=\log g(x)$ and $h'(x)$, the first differential of $h(\cdot)$ has been evaluated
at $m$ abscissae in $D:x_1\leq x_2\leq\cdots x_m$. For $j=1,\ldots,m-1$, define
\begin{equation}
v_j=\frac{h(x_{j+1})-h(x_j)-x_{j+1}h'(x_{j+1})+x_jh'(x_j)}{h'(x_j)-h'(x_{j+1})}
\label{eq:v}
\end{equation}
For $x\in[v_{j-1},v_j]$; $j=1,\ldots,m$, define
\begin{equation}
u_m(x)=h(x_j)+(x-x_j)h'(x_j),
\label{eq:u}
\end{equation}
Here $v_0$ is the lower bound of $D$ (or $-\infty$ if $D$ is not bounded below)
and $v_m$ is the upper bound of $D$ (or $\infty$ if $D$ is not bounded above).
Also define
\begin{equation}
s_m(x)=\frac{\exp\{u_m(x)\}}{\int_D\exp\{u_m(x')\}dx'}
\label{eq:s}
\end{equation}
Also define, for $x\in[x_j,x_{j+1}]$; $j=1,\ldots,m-1$,
\begin{equation}
\ell_m(x)=\frac{(x_{j+1}-x)h(x_j)+(x-x_j)h(x_{j+1})}{x_{j+1}-x_j},
\label{eq:ell}
\end{equation}
and for $x<x_1$ or $x>x_m$, $l_m(x)=-\infty$. Thus, for all
$x\in D$, we have, due to concavity of $h(\cdot)$,
\begin{equation}
\ell_m(x)\leq h(x)\leq u_m(x)
\label{eq:squeeze}
\end{equation}

To sample using ARS, draw $x^*\sim s_m$ and $w\sim Uniform(0,1)$ independently
and accept $x^*$ if $w\leq \exp\{\ell_m(x^*)-u_m(x^*)\}$. Else accept $x^*$
if $w\leq \exp\{h(x^*)-u_m(x^*)\}$. If the sampling is to be continued then the
accepted values may be included in the set of abscissae (the latter to be re-arranged
in ascending order) to adaptively make the bounds (\ref{eq:squeeze}) finer and finer;
this enhances efficiency as the sampling progresses.

For our purpose of perfect sampling using the log-concave full conditionals
we shall need to represent the Gibbs transition kernel in a special form using the lower bound of the form
given in (\ref{eq:squeeze}), while using ARS in conjunction for sampling.
We introduce our perfect sampling idea in the next section.

\subsection{Construction of perfect simulation methodology in posteriors with log-concave full conditionals}
\label{subsec:perfect_logconcave}

For the sake of generality, we consider full conditionals of the form
$\pi_i(\xi_i)=\pi(\xi_i\mid \bxi_{-i});i=1,\ldots,p$, where it is necessary to simulate
perfectly from the joint distribution of $\bxi=\{\xi_1,\ldots,\xi_p\}$;
here $\bxi_{-i}=\bxi\backslash \xi_i$. We
assume that each $\pi_i(\xi_i)$ is log-concave.
It then follows from (\ref{eq:squeeze}) that
\begin{equation}
\pi_i(\xi)\geq\exp\{\ell_{m,\bxi_{-i}}(\xi)\},
\label{eq:lb1}
\end{equation}
where $\ell_{m,\bxi_{-i}}(\cdot)$ may depend upon $\bxi_{-i}$. Taking the infimum over $\bxi_{-i}$ yields
\begin{equation}
\pi_i(\xi)\geq\exp\{\ell_{m,\bxi_{-i}}(\xi)\}\geq\inf_{\bxi_{-i}}\exp\{\ell_{m,\bxi_{-i}}(\xi)\}=\exp\{\ell_{m,i}(\xi)\},
\label{eq:lb2}
\end{equation}
where $\exp\{\ell_{m,i}(\xi)\}=\inf_{\bxi_{-i}}\exp\{\ell_{m,\bxi_{-i}}(\xi)\}$ is independent of $\bxi_{-i}$.
However, the right hand side of (\ref{eq:lb2}) need not be a density in that it need not integrate to one.
Firstly, finiteness of the integral can be ensured
at least if $\bxi$ is restricted to a compact set. That restriction of the support of the
parameters to some judicously constructed compact set is not unrealistic is discussed in detail in \ctn{Sabya10b}.
Let $D_i$ denote a compact interval to which $\xi_i$ is restricted. Let $\epsilon_i=\int_{D_i}\exp\{\ell_{m,i}(\xi)\}$,
and let $g_{m,i}(\xi)=\epsilon^{-1}_i\exp\{\ell_{m,i}(\xi)\}$ denote the density corresponding to $\exp\{\ell_{m,i}(\xi)\}$.
Then, we have, for each $i=1,\ldots,p$,
\begin{equation}
\pi_i(\xi)\geq\epsilon_ig_{m,i}(\xi),
\label{eq:lb3}
\end{equation}
which implies that
\begin{equation}
\prod_{i=1}^p\pi_i(\xi_i)\geq\prod_{i=1}^p\epsilon_ig_{m,i}(\xi_i)=\epsilon g_m(\bxi),
\label{eq:lb4}
\end{equation}
where $\epsilon=\prod_{i=1}^p\epsilon_i$ and $g_m(\bxi)=\prod_{i=1}^pg_{m,i}(\xi_i)$.
That $0<\epsilon\leq 1$ is clear since for each $i=1,\ldots,p$, $0<\epsilon_i\leq 1$, the latter following
by integrating both sides of (\ref{eq:lb3}) over the support $D_i$.

This then implies that the Gibbs transition kernel, given by
\begin{equation}
P(\bxi^{(t+1)}\mid\bxi^{(t)})=\prod_{i=1}^p
\pi(\xi^{(t+1)}_i\mid\xi^{(t+1)}_1,\ldots,\xi^{(t+1)}_{i-1},\xi^{(t)}_{i+1},\ldots,\xi^{(t)}_p),
\label{eq:gibbs_kernel}
\end{equation}
can be represented as
\begin{equation}
P(\bxi^{(t+1)}\mid\bxi^{(t)})=\epsilon g_m(\bxi^{(t+1)})+(1-\epsilon)R_m(\bxi^{(t+1)}\mid\bxi^{(t)}),
\label{eq:rep1}
\end{equation}
where
\begin{equation}
R_m(\bxi^{(t+1)}\mid\bxi^{(t)})=\frac{P(\bxi^{(t+1)}\mid\bxi^{(t)})-\epsilon g_m(\bxi^{(t+1)})}{1-\epsilon}
\label{eq:residual}
\end{equation}
is the residual density.

Hence, in order to simulate the Gibbs chain $P(\bxi^{(t+1)}\mid\bxi^{(t)})$ one can first draw $\delta^{(t+1)}\sim Bernoulli(\epsilon)$;
if $\delta^{(t+1)}=1$, then $\bxi^{(t+1)}$ is drawn from $g_m(\cdot)$, and if $\delta^{(t+1)}=0$, $\bxi^{(t+1)}\sim R_m(\cdot\mid\bxi^{(t)})$.
Thus, if $\delta^{(t+1)}=1$, then $\bxi^{(t+1)}$ is drawn from $g_m(\cdot)$ which does not depend upon the previous iteration $\bxi^{(t)}$.
We shall exploit this fact for our perfect sampling algorithm. Indeed, we constructed the mixture representation (\ref{eq:rep1})
just to achieve this independence of $\bxi^{(t)}$ which happens with positive probability $\epsilon$. This implies that once
$\delta^{(t)}=1$ for some $t<0$ in the associated coupling from the past algorithm (CFTP), all possible chains initialised
at all possible values of the state-space, will coalesce!

But since drawing directly from $R_m(\cdot\mid\bxi^{(t)})$ (necessary when $\delta^{(t)}=0$) is not straightforward, we consider
a rejection sampling method which we now describe. Note that
\begin{align}
R_m(\bxi^{(t+1)}\mid\bxi^{(t)})&=\frac{P(\bxi^{(t+1)}\mid\bxi^{(t)})-\epsilon g_m(\bxi^{(t+1)})}{1-\epsilon}\label{eq:rs1}\\
&\leq \frac{P(\bxi^{(t+1)}\mid\bxi^{(t)})}{1-\epsilon}\label{eq:rs2}
\end{align}
Hence, we consider the rejection sampling scheme provided in Algorithm \ref{algo:rs1}.
\begin{algo}\label{algo:rs1}\topline
Rejection sampling from $R_m(\cdot\mid\bxi^{(t)})$ \botline \normalfont \ttfamily
\begin{itemize}
\item[(1)] Draw $\bxi\sim P(\cdot\mid\bxi^{(t)})$ by sampling from the full conditionals,
and independently draw $U\sim Uniform(0,1)$.
\item[(2)] Accept $\bxi$ as a realization from $R_m(\cdot\mid\bxi^{(t)})$ if
$$U<(1-\epsilon)\frac{R_m(\bxi\mid\bxi^{(t)})}{P(\bxi\mid\bxi^{(t)})}.$$
\end{itemize}
\rmfamily
\botline
\end{algo}
Note that sampling from $P(\cdot\mid\bxi^{(t)})$ may require ARS from the full conditionals.
To avoid ARS one may further upper bound $P(\bxi\mid\bxi^{(t)})$ using the upper bounds
available for log-concave densities as follows.
\begin{equation}
P(\bxi\mid\bxi^{(t)})\leq\eta f_m(\bxi),
\label{eq:ub1}
\end{equation}
where $\eta=\prod_{i=1}^p\eta_i$,
$\eta_i=\int_{D_i}\exp\{u_{m,i}(\xi)\}d\xi$,
$u_{m,i}(\xi)=\sup_{\bxi_{-i}}u_{m,\bxi_{-i}}(\xi)$,
and
\begin{equation}
f_m(\bxi)=\prod_{i=1}^pf_{m,i}(\xi_i),
\label{eq:ub2}
\end{equation}
with $f_{m,i}(\xi)=\eta^{-1}_i\exp\{u_{m,i}(\xi)\}$.
Since we also have the lower bound $P(\bxi\mid\bxi^{(t)})\geq\epsilon g_m(\bxi)$, the following
rejection sampling method given by Algorithm (\ref{algo:rs2}) can be employed to sample from $P(\cdot\mid\bxi^{(t)})$.
\begin{algo}\label{algo:rs2}\topline
Rejection sampling from $P(\cdot\mid\bxi^{(t)})$ \botline \normalfont \ttfamily
\begin{itemize}
\item[(1)] Draw $\bxi\sim f_m(\cdot)$,
and independently draw $U\sim Uniform(0,1)$.
\item[(2)] Accept $\bxi$ as a realization from $P(\cdot\mid\bxi^{(t)})$ if
$$U<\frac{\epsilon g_m(\bxi)}{\eta f_m(\bxi)}.$$
\item[(3)] Else accept $\bxi$ as a realization from $P(\cdot\mid\bxi^{(t)})$ if
$$U<\frac{P(\bxi\mid\bxi^{(t)})}{\eta f_m(\bxi)}.$$
\end{itemize}
\rmfamily
\botline
\end{algo}
It is important to remark that whenever simulation from $P(\cdot\mid\bxi^{(t)})$ is straightforward,
that is, whenever the full conditionals $\pi_i(\xi_i)$ are of standard forms, rejection sampling
or ARS will not be used for sampling from the Gibbs kernel.

Our mixture Gibbs kernel (\ref{eq:rep1}) resembles that associated with the ``multigamma coupler"
of \ctn{Murdoch98}, but the latter is a representation of one-dimensional cases only. Moreover, such
mixture representation is very rarely achievable in reality for densities that are not log-concave. Perfect
simulation of high-dimensional variables, using the one-dimensional multigamma coupler for each univariate full conditional
densitiy, is possible in principle, but is likely to be extremely inefficient, particularly as
the dimension of the random variable tends to be large.
As is evident from our construction, we completely
bypass such difficulties by representing the Gibbs kernel of the entire high-dimensional random variable
$\bxi$ as a mixture of two (high-dimensional) densities, obtained using properties of log-concavity
of the full conditionals. We have also shown how to sample from the two high-dimensional densities.
In particular, we have provided an explicit rejection sampling method for simulating from the
residual density of the mixture representation, whatever the dimensionality. We remark that explicit
methods of simulating from the residual density has not been provided in \ctn{Murdoch98}
or \ctn{Green99}. Although \ctn{Mykland95} proposed a trick to completely avoid simulation
from the residual density in the context of regenerative simulation, such trick is not
applicable in perfect simulation.

For perfect simulation we exploit the following idea first presented in \ctn{Murdoch98}.
Note that there is a fixed probability $\epsilon$ that at any given time $T=t$, $\bxi$ will be drawn
from $g_m(\cdot)$. Hence, $T$ follows a geometric distribution given by
$P(T=t)=\epsilon (1-\epsilon)^t$; $t=0,1,2,\ldots$. As a result, it is possible to simulate
$T$ from the geometric distribution and then draw $\bxi^{(-T)}\sim g_m(\cdot)$. Then the chain only
need to be carried forward in time till time $t=0$, using $\bxi^{(t+1)}=\psi(\bxi^{(t)},\bU^{(t+1)})$,
where $\psi(\bxi^{(t)},\bU^{(t+1)})$ is the deterministic function corresponding to the simulation
of $\bxi^{(t+1)}$ from $R_m(\cdot\mid\bxi^{(t)})$ using the set of appropriate random numbers $\bU^{(t+1)}$;
the sequence $\{\bU^{(t)};t=0,-1,-2,\ldots\}$ being assumed to be available before beginning the perfect
sampling simulation. The resulting draw $\bxi^{(0)}$ sampled at time $t=0$ is a perfect sample
from $\pi(\bxi)$. For subsequent reference we present this in an algorithmic way in Algorithm \ref{algo:perfect_logconcave}.
\begin{algo}\label{algo:perfect_logconcave}\topline
Perfect simulation from $\pi(\bxi)$ \botline \normalfont \ttfamily
\begin{itemize}
\item[(1)] Draw $T\sim Geometric(\epsilon)$.
\item[(2)] Draw $\bxi^{(-T)}\sim g_m(\cdot)$.
\item[(3)] Carry the chain forward till time $t=0$ using the deterministic functional
relationship $\bxi^{(t+1)}=\psi(\bxi^{(t)},\bU^{(t+1)})$ and the available sequence
$\{\bU^{(t)};t=0,1,2,\ldots\}$.
\item[(4)] Report $\bxi^{(0)}$ as a perfect sample from $\pi(\bxi)$.
\end{itemize}
\rmfamily
\botline
\end{algo}
The above perfect sampling algorithm is to be embedded in the perfect sampling algorithm
for mixture simulation in the context of categorical time series. This we do in the next section.

\section{Perfect simulation for mixtures of categorical time series with unknown number of components}
\label{sec:perfect_categorical}

We first note that coalescence of $(Z,C,S)$ (equivalently, coalescence of  $(Z,C)$ since
coalescence of $C$ implies coalescence of $S$) implies coalescence of $(\bPhi,\bgamma)$.
We exploit the bounding chains construction approach of \ctn{Sabya10b} for facilitating coalescence.
The idea is to obtain stochastic lower and upper bounds for the discrete parts of the Gibbs sampler, namely
for $(Z,C,S)$, by maximizing and minimizing their respective distribution functions
with respect to the continuous parameters, simulating only from the lower and the upper bounding chains thus
created, and noting their coalescence.
Remarkably, there is no need to simulate the continuous parameters $\bPhi$ and $\bgamma$ before coalescence
of $(Z,C,S)$. Simulation of $(\bPhi,\bgamma)$, conditional on $(Z,S)$, is necessary only after the coalescence of
the latter. However, as already discussed in Section \ref{subsec:perfect_need},
methods for directly simulating $(\bPhi,\bgamma)$ given $(Z,S)$ are not available, and we must resort
to the perfect simulation method introduced in Section \ref{sec:perfect_logconcave} using the available
full conditionals which are, thankfully, log-concave.

We now proceed to construction of appropriate bounding chains for the discrete parameters $(Z,C,S)$.

\subsection{Bounding chains}
\label{subsec:sb_bounding_chains}

\subsubsection{Bounds for $Z$}
\label{subsubsec:z_bound}

Let $F_{z_i}(\cdot\mid \bY,S,k,\bTheta_M)$ denote
the distribution function of the full conditional of $z_i$, and let $F_{c_j}(\cdot\mid \bY, S_{-j},k_j,\bPhi)$ and
$F_{s_j}(\cdot\mid \bY, S_{-j},C,\bTheta_M)$ stand for those of $c_j$ and $s_j$, respectively.
In addition, when required, we shall assume that 
$\{\gamma_{st}\}$ have compact supports not containing zero. This assumption entails multiplication
of a constant to the prior to take care of the truncation, but clearly this constant does not destroy
the log-concavity of the full conditional of $\gamma_{st}$. On the other hand, truncation of $\phi_{\ell,st}$
would involve a factor that depends upon $\gamma_{st}$, which might affect log-concavity of $\gamma_{st}$.
However, we did not find truncation of $\phi_{\ell,st}$ to be necessary in our simulations.

Letting $\bar S$ denote the set consisting of only those $s_j$ that have coalesced, and let $S^-=S\backslash\bar S$ consist of the
remaining $s_j$. Then
\begin{eqnarray}
F^L_{z_i}\left(\cdot\mid \bY,\bar S\right)&=&\inf_{S^-,k,\bPhi}F_{z_i}(\cdot\mid \bY,\bar S,S^-,k,\bPhi)\label{eq:inf_z}\\
F^U_{z_i}\left(\cdot\mid \bY,\bar S\right)&=&\sup_{S^-,k,\bPhi}F_{z_i}(\cdot\mid \bY,\bar S,S^-,k,\bPhi)\label{eq:sup_z}
\end{eqnarray}
Fixing $\bar S$ helps reduce the gap between (\ref{eq:inf_z}) and (\ref{eq:sup_z}).
As in \ctn{Sabya10b} we calculate the infimum and the supremum above by simulated annealing.
For further details, wee \ctn{Sabya10b}.


\subsubsection{Bounds for $C$}
\label{subsubsec:c_bound}
Let $\bar Z$ denote the set of coalesced $z_i$, and let $Z^-=Z\backslash\bar Z$ consist of those $z_j$ that did not yet coalesce.
Then
\begin{eqnarray}
F^L_{c_j}\left(\cdot\mid \bY,\bar S, \bar Z\right)&=&\inf_{S^-,k_j,Z^-,\bPhi}F_{c_j}(\cdot\mid \bY,\bar S,S^-,k_j,\bar Z,Z^-,\bPhi)\label{eq:inf_c}\\
F^U_{c_j}\left(\cdot\mid \bY,\bar S, \bar Z\right)&=&\sup_{S^-,k_j,Z^-,\bPhi}F_{c_j}(\cdot\mid \bY,\bar S,S^-,k_j,\bar Z,Z^-,\bPhi)\label{eq:sup_c}
\end{eqnarray}
As noted in \ctn{Sabya10b}, the supremum corresponds to $k_j=1$ and the infimum corresponds to $k_j=M-1$.
For details on optimization using simulated annealing, see \ctn{Sabya10b}.


\subsubsection{Bounds for $S$}
\label{subsubsec:s_bound}

Letting $\bar C$ and $C^-=C\backslash\bar C$ denote the sets of coalesced and the non-coalesced $c_j$, the lower
and the upper bounds for the distribution function of $s_j$ are
\begin{eqnarray}
F^L_{s_j}\left(\cdot\mid \bY,\bar C\right)&=&\inf_{C^-,\bPhi}F_{s_j}(\cdot\mid \bY,\bar C,C^-,\bPhi)\label{eq:inf_s}\\
F^U_{s_j}\left(\cdot\mid \bY,\bar C \right)&=&\sup_{C^-,\bPhi}F_{s_j}(\cdot\mid \bY,\bar C,C^-,\bPhi)\label{eq:sup_s}
\end{eqnarray}
Optimization in this case requires careful attention; see \ctn{Sabya10b} for details.
\begin{algo}\label{algo:cftp_unknown}\topline
CFTP for mixtures with unknown number of components \botline \normalfont \ttfamily
\begin{itemize}
\item[(1)] For $j=1\ldots$, until coalescence of $(Z,C)$, repeat steps (2) and (3) below.
\item[(2)] Define $\mathcal S_j=\{-2^j+1,\ldots,-2^{j-1}\}$ for $j\geq 2$,
and let $\mathcal S_1=\{-1,0\}$. 
For each $m\in\mathcal S_j$,
generate random numbers $R_{Z,m}$, $R_{C,m}$, $R_{S,m}$, $R_{\bTheta_M,m}$, and $R_{\bgamma,m}$, meant
for simulating $Z$, $C$, $S$, $\bTheta_M$, and $\bgamma$ respectively.
Note that for each $m$, $R_{\bTheta,m}$ and $R_{\bgamma,m}$ are random numbers
corresponding to the perfect simulation algorithm given by Algorithm \ref{algo:perfect_logconcave};
$(\bPhi,\bgamma)$ in this problem corresponds to the random vector $\bxi$ in that algorithm.
We need to generate and fix these random numbers even though we won't actually simulate $(\bPhi,\bgamma)$
before coalescence of $(Z,C)$.

Once generated, treat the random numbers as fixed thereafter for all iterations.
Since step $-2^j$ is the initializing step, no random number generation is required
at this step.
\item[(3)] For $t=-2^j+1,\ldots,-1,0$, implement steps (3) (i), (3) (ii) and (3) (iii):
\begin{itemize}
\item[(i)] For $i=1,\ldots,n$,
\begin{itemize}
\item[(a)] For $\ell=1,\ldots,M$, calculate $F^L_{z_i}(\ell\mid \bY,\bar S)$ and
$F^U_{z_i}(\ell\mid \bY,\bar S)$ using the simulated annealing techniques detailed
in \ctn{Sabya10b}.
\item[(b)] Determine $z^L_{it}=F^{U-}_{z_i}(R_{z{i,t}}\mid \bY,\bar S)$ and
$z^U_{it}=F^{L-}_{z_i}(R_{z{i,t}}\mid \bY,\bar S)$.
\end{itemize}
\item[(ii)] For $i=1,\ldots,M$,
\begin{itemize}
\item[(a)] For $\ell=1,\ldots,k_i+1$, calculate $F^L_{c_i}(\ell\mid \bY,\bar S,\bar Z)$ and
$F^U_{c_i}(\ell\mid \bY,\bar S,\bar Z)$, using the simulated annealing techniques
of \ctn{Sabya10b}. The supremum corresponds to $k_i=\#\bar S\backslash\{s_i\}$, when $S^-$ contains
a single distinct element, and the infimum corresponds to the case where
$k_i=\# \left(\bar S\cup S^-\right)\backslash\{s_i\}$, when all elements of $S^-$ are distinct, and so
the set $S^-$ will be set manually to have a single distinct element or all distinct elements.

\item[(b)] Set $c^L_{it}=F^{U-}_{c_i}(R_{c_{i,t}}\mid \bY,\bar S,\bar Z)$ and
$c^L_{it}=F^{U-}_{c_i}(R_{c_{i,t}}\mid \bY,\bar S,\bar Z)$.
\end{itemize}

\item[(iii)] For $i=1,\ldots,M$,
\begin{itemize}
\item[(a)] For $\ell=1,\ldots,M$, calculate $F^L_{s_i}(\ell\mid \bY,\bar C)$ and
$F^U_{s_i}(\ell\mid \bY,\bar \bC)$, using simulated annealing techniques detailed in \ctn{Sabya10b}.

\item[(b)]
Since, for some $\ell^*\in\{1,\ldots,M\}$,
$F^L_{s_i}(\ell\mid \bY,\bar C)=0$ for $\ell<\ell^*$ and 1 for $\ell\geq \ell^*$,
it follows that $s^L_{it}=\ell^*$. Similarly, $s^U_{it}$ can be determined.
\end{itemize}

\end{itemize}
\item[(4)] If, for some $t^*<0$, $z^L_{it^*}=z^U_{it^*}$ $\forall i=1,\ldots,n$, and $c^L_{it^*}=c^U_{it^*}$
$\forall i=1,\ldots,M$, then
run the following Gibbs sampling steps from $t=t^*$ to $t=0$:
\begin{itemize}
\item[(a)] Let $Z^*=\{z^*_1,\ldots,z^*_n\}$ and $C^*=\{c^*_1,\ldots,c^*_M\}$ denote the coalesced values of $Z$
and $C$ respectively, at time $t^*$.
Given $(Z^*,C^*)$, arbitrarily choose any value of $\bTheta_M$ which is compatible with $C^*$
(one way to ensure compatibility is to choose any $\bTheta_M$ having $M$ distinct elements); then
obtain $S^*$ from $[S\mid\bY,C,\bTheta_M]$ using the algorithm given in Section \ref{subsec:relabeling}.

\item[(b)] Finally, generate $(\bPhi,\bgamma)$ using the perfect simulation algorithm
described in Algorithm \ref{algo:perfect_logconcave}, using the random numbers
already generated.
This yields the coalesced value $(Z^*,C^*,S^*,\bPhi^*,\bgamma^*)$ at time $t=t^*$.

\item[(b)] Using the random numbers already generated, carry forward the above Gibbs sampling
chain started at $t=t^*$ till $t=0$,
simulating, in order, from the full conditionals of the individual components of $(Z,C,S)$, provided in
Sections \ref{subsec:fullcond_z}, \ref{subsec:fullcond_c}, and \ref{subsec:fullcond_s}, and by perfectly simulating
$(\bPhi,\bgamma)$ using Algorithm \ref{algo:perfect_logconcave}. Note that, once $(\bPhi^*,\bgamma^*)$
are generated by perfect sampling at time $t=t^*$, further perfect sampling of $(\bPhi,\bgamma)$ for time $t>t^*$
does not seem necessary since now Gibbs sampling can be employed. But somewhat ironically, we are forced
to continue perfect sampling since changing the simulation method in the midway is not legitimate.

\item[(c)]Then, the output of the Gibbs sampler obtained at $t=0$,
which we denote by $(Z_0,C_0,S_0,\bPhi_0,\bgamma_0)$, is a perfect sample
from the true target posterior distribution.

\end{itemize}
\end{itemize}
\rmfamily
\botline
\end{algo}

\renewcommand\thefigure{S-\arabic{figure}}
\renewcommand\thetable{S-\arabic{table}}
\renewcommand\thesection{S-\arabic{section}}

\begin{center}
{\bf\LARGE Supplementary Material}
\end{center}


\section{Marginalized full conditional distributions}
\label{sec:marginalized_fullcond}


\subsection{Marginalized full conditionals of $\{z_1,\ldots,z_n\}$}
\label{subsec:marginalized_fullcond_z}

For $i^*=1,\ldots,n$, let $Z_{i^*}=\{z_1,\ldots,z_{i^*-1},z_{i^*+1},\ldots,z_n\}$, and
let $C$ consist of $k$ distinct components. Then, denoting the set $\{\gamma_{st};s,t=1,\ldots,K\}$
by $\bgamma$, the full conditional distribution of $z_{i^*}$ is given by
\begin{align}
[z_{i^*}=r\mid Z_{-i^*},C,\bgamma,k]&\propto \prod_{\ell=1}^k\prod_{s=1}^K
\frac{\prod_{t=1}^K\Gamma\left(\sum_{i:z_i=j}\sum_{j:c_j=\ell}N_{i,st}+\gamma_{st}\right)}
{\Gamma\left(\sum_{t=1}^K\sum_{i:z_i=j}\sum_{j:c_j=\ell}N_{i,st}+\sum_{t=1}^K\gamma_{st}\right)}
\label{eq:marginalized_fullcond_z}
\end{align}
In the right hand side of (\ref{eq:fullcond_z}), $z_{i^*}$ must be replaced with $r$.

\subsection{Marginalized full conditionals of $\{c_1,\ldots,c_M\}$}
\label{subsec:marginalized_fullcond_c}

To obtain the full conditional of $c_r;r=1,\ldots,M$, first let $k_{r}$ denote the number of distinct values in 
$\bTheta_{-rM}=\{\btheta_1,\ldots,\btheta_{r-1},\btheta_{r+1},\ldots,\btheta_M\}$, and let 
$\bphi^{(r^*)}_{\ell}$; $\ell=1,\ldots,k_{r}$ denote
the distinct values. 
Also suppose that $\bphi^{(r^*)}_{\ell}$ occurs $M_{\ell r}$ times.
Then the conditional distribution of $c_r$ is given by
\begin{equation}
[c_r=\ell\mid Y,Z,C_{-r},\bgamma,k_r]=\left\{\begin{array}{c}\kappa q_{\ell r}\hspace{2mm}\mbox{if}\hspace{2mm}
\ell=1,\ldots,k_r\\ \kappa q_{0r}\hspace{2mm}\mbox{if}\hspace{2mm}\ell=k_r+1\end{array}\right.
\label{eq:marginalized_fullcond_c}
\end{equation}
where
\begin{align}
q_{\ell r}&\propto M_{\ell r}\times
\prod_{s=1}^K
\frac{\Gamma\left(\sum_{t=1}^K\sum_{i:z_i=j,j:c_j=\ell,j\neq r}N_{i,st}+\sum_{t=1}^K\gamma_{st}\right)}
{\prod_{t=1}^K\Gamma\left(\sum_{i:z_i=j,j:c_j=\ell,j\neq r}N_{i,st}+\gamma_{st}\right)}\notag\\
&\times \prod_{s=1}^K
\frac{\prod_{t=1}^K\Gamma\left(\sum_{i:z_i=r}N_{i,st}+\sum_{i:z_i=j,j:c_j=\ell,j\neq r}N_{i,st}+\gamma_{st}\right)}
{\Gamma\left(\sum_{t=1}^K\sum_{i:z_i=r}N_{i,st}+\sum_{t=1}^K\sum_{i:z_i=j,j:c_j=\ell,j\neq r}N_{i,st}+\sum_{t=1}^K\gamma_{st}\right)}
\label{eq:marginalized_fullcond_c1}
\end{align}
and
\begin{align}
q_{0 r}&\propto \alpha\times
\prod_{s=1}^K
\frac{\Gamma\left(\sum_{t=1}^K\gamma_{st}\right)}
{\prod_{t=1}^K\Gamma\left(\gamma_{st}\right)}
\times\prod_{s=1}^K
\frac{\prod_{t=1}^K\Gamma\left(\sum_{i:z_i=r}N_{i,st}+\gamma_{st}\right)}
{\Gamma\left(\sum_{t=1}^K\sum_{i:z_i=r}N_{i,st}+\sum_{t=1}^K\gamma_{st}\right)}
\label{eq:marginalized_fullcond_c2}
\end{align}

\subsection{Marginalized full conditionals of $\{\gamma_{st};s,t=1,\ldots,K\}$}
\label{subsec:marginalized_fullcond_gamma}

Assuming that $C$ consists of $k$ distinct components, the full
conditional distribution of $\gamma_{s^*t^*}$, for 
$s^*=1,\ldots,K$, and $t^*=1,\ldots,K$, is given by
\\[2mm]
$[\gamma_{\ell,s^*t^*}\mid Z,C,\bgamma_{-s^*,-t^*},k]$ 
\begin{align}
&\propto 
\left\{\prod_{\ell=1}^k\frac{\Gamma\left(\sum_{i:z_i=j}\sum_{j:c_j=\ell}N_{i,s^*t^*}+\gamma_{s^*t^*}\right)}
{\Gamma\left(\sum_{t=1}^K\sum_{i:z_i=j}\sum_{j:c_j=\ell}N_{i,s^*t}+\sum_{t=1}^K\gamma_{s^*t}\right)}\right\}
\times 
\left(\frac{\Gamma\left(\sum_{t=1}^K\gamma_{s^*t}\right)}
{\Gamma\left(\gamma_{s^*t^*}\right)}\right)^k\notag\\
&\times \gamma^{a_{jk}-1}_{s^*t^*}\exp\left(-b_{jk}\gamma_{s^*t^*}\right)\notag\\
&= \prod_{\ell=1}^k\left\{\frac{\Gamma\left(\sum_{i:z_i=j}\sum_{j:c_j=\ell}N_{i,s^*t^*}+\gamma_{s^*t^*}\right)}
{\Gamma\left(\sum_{t=1}^K\sum_{i:z_i=j}\sum_{j:c_j=\ell}N_{i,s^*t}+\sum_{t=1}^K\gamma_{s^*t}\right)}
\times 
\frac{\Gamma\left(\sum_{t=1}^K\gamma_{s^*t}\right)}
{\Gamma\left(\gamma_{s^*t^*}\right)}\right\}\label{eq:gamma1}\\
&\times \gamma^{a_{jk}-1}_{s^*t^*}\exp\left(-b_{jk}\gamma_{s^*t^*}\right)\notag
\end{align}

\subsection{Discussion on log-concavity of the full conditionals of $\{\gamma_{st};s,t=1,\ldots,K\}$}
\label{subsec:gamma_logconcavity}

Note that each factor in the product (\ref{eq:gamma1}) is of the form
\begin{equation}
h(\gamma_{s^*t^*})=\frac{\Gamma(\gamma_{s*t^*}+a_{t^*})}{\Gamma(\gamma_{s^*t^*})}
\times\frac{\Gamma(\gamma_{s^*t^*}+y_{\ell,t^*})}{\Gamma(\gamma_{s^*t^*}+y_{\ell,t^*}+a_{t^*}+b_{\ell,t^*})},
\label{eq:gamma2}
\end{equation}
$a_{t^*}=\sum_{t=1,t\neq t^*}^K\gamma_{s^*t}$, $y_{\ell,t}=\sum_{i:z_i=j}\sum_{j:c_j=\ell}N_{i,s*t}$ $\forall t=1,\ldots,K$,
and $b_{\ell,t^*}=\sum_{t=1,t\neq t^*}^Ky_{\ell,t}$. Clearly, all the terms are non-negative, with $y_{\ell,t}$ and $b_{\ell,t^*}$
being integers.
As a result, $h(\gamma_{s^*t^*})$ admits the following simple form:
\begin{equation}
h(\gamma_{s^*t^*})=\frac{\prod_{i=1}^{y_{t^*}}(\gamma_{s^*t^*}+y_{t^*}-i)}{\prod_{i=1}^{y_{t^*}+b_{\ell,t^*}}
(\gamma_{s^*t^*}+a_{t^*}+y_{t^*}+b_{\ell,t^*}-i)}.
\label{eq:gamma3}
\end{equation}
Thus,
\begin{align}
\frac{d^2\log h(\gamma_{s^*t^*})}{d\gamma^2_{s^*t^*}}
&=-\sum_{i=1}^{y_{t^*}}\left\{\frac{1}{\left(\gamma_{s^*t^*}+y_{t^*}-i\right)^2}
-\frac{1}{\left(\gamma_{s^*t^*}+a_{t^*}+y_{t^*}-i\right)^2}\right\}\notag\\
& \ \ \ \ +\sum_{i=y_{t^*}+1}^{y_{t^*}+b_{\ell,t^*}}\frac{1}{\left(\gamma_{s^*t^*}+a_{t^*}+y_{t^*}-i\right)^2}.
\label{eq:gamma5}
\end{align}
Unless $b_{\ell,t^*}=0$ (that is, $y_{\ell,t}=0$ $\forall t\neq t^*$), (\ref{eq:gamma5}) need not be negative for all 
$\gamma_{s^*t^*}$ and $\bgamma_{-s^*,-t^*}$.


\bibliography{irmcmc}

\end{document}